# Nonlocality Versus Modified Realism

**Hervé Zwirn**

CMLA (ENS Paris Saclay, 61 avenue du Président Wilson 94235 Cachan, France)
&
IHPST (CNRS, ENS Ulm, University Paris 1, 13 rue du Four, 75006 Paris, France)
herve.zwirn@gmail.com

**Abstract:** A large number of physicists now admit that quantum mechanics is a non-local theory. The EPR argument and the many experiments (including recent "loop-hole free" tests) showing the violation of Bell's inequalities seem to have confirmed convincingly that quantum mechanics cannot be local. Nevertheless, this conclusion can only be drawn inside a standard realist framework assuming an ontic interpretation of the wave function and viewing the collapse of the wave function as a real change of the physical state of the system. We show that this standpoint is not mandatory and that if the collapse is not considered an actual physical change it is possible to recover locality.

1. **Introduction**

Quantum mechanics, though the most efficient and precise theory of physics, is still open to numberless debates about the correct interpretation of its formalism almost one century after its inception. It is most of the time associated to superposition, indeterminism, entanglement, contextuality and non-locality, all strange features which go away from the world classical physics has accustomed us to.

Deciding if the world is local or not is a very exciting question because it is both of empirical and epistemological nature. It is usually held that the empirically verified violation of Bell's inequalities [1] implies that non-locality is unavoidable. But this conclusion depends on the widely debated question of the status conferred to the quantum states. Are they corresponding directly to reality or are they representing only our knowledge of some aspect of reality? This is the now classical distinction between the ontic and the epistemic interpretations of the wave function [2]. The ontic view posits that quantum states refer directly to reality. The epistemic view on the contrary assumes that quantum states refer only to our knowledge of part of reality. Moreover, the quantum state can be considered as a complete description of reality - which is called the $\psi$-complete view - or may require to be supplemented with additional variables –this is the $\psi$-supplemented view [3]. The $\psi$-complete view is ontic and constitutes the orthodox interpretation of Quantum Mechanics. In the $\psi$-supplemented view, it is assumed that the real state of the system, usually denoted by $\lambda$, is not totally specified by the quantum state which is compatible with several $\lambda$. So there is a probability distribution of $\lambda$ over $\psi$, $p(\lambda|\psi)$. Harrigan and Spekkens [3] noticed that if the probability distributions associated with two different quantum states $\psi$ and $\psi'$ are never overlapping, then each ontic state $\lambda$ encodes only one quantum state because one $\lambda$ is consistent with only one $\psi$. In this sense the quantum state can be considered as ontic because a variation of ψ implies a variation of reality and so, even incomplete, ψ captures a true aspect of reality. On the contrary, if the probability distributions associated with two different quantum states ψ and ψ' are overlapping, then ψ and ψ' are compatible at least with one ontic state. Then they can be considered as referring simply to our knowledge much in the same way as probability distributions in statistical mechanics are. This is the ψ-epistemic view.

Many results have been proved for the models fitting this taxonomy. Under the assumption that systems that are prepared independently have independent physical states, Pusey et al. [4] show that if the quantum state can be specified more precisely than the ψ description, (let λ be a complete description of the physical state and let one ψ corresponds to several λ) then one λ cannot belong to several ψ. That means that the ψ-epistemic view is ruled out. A stronger claim is made by Colbeck and Renner [5] who prove that under the assumptions that measurement settings can be chosen freely and that there is no superluminal signalling of the choice at the ontic level, any model must be ψ-complete. It seems then that under reasonable enough assumptions the ψ-epistemic view is difficult to support. Nevertheless, Lewis et al. [6] show that if the preparation independence assumption is abandoned and if superluminal influences of measurement choices upon ontic variables are allowed then it is possible to build a ψ-epistemic model.

It is true that if we consider that pure quantum states refer directly to real states of the system (whether or not completed by hidden variables), non-locality seems difficult to avoid [3]. Harrigan and Spekkens argue that it is possible to show that locality is ruled out for ψ-ontic models even without the need of Bell's inequalities. But if we are confident that the violation of Bell's inequalities (or equivalent forms) has been proved now without any doubt, as many experiments (including recent "loop-hole free" tests) [7, 8, 9, 10, 11, 12, 13] seem to show, it seems difficult to escape the conclusion that the world is not local.

Now Spekkens's taxonomy is relevant mainly in a realist context such that there is "really" a physical system independent of the presence or of the knowledge of any observer and that to this system corresponds an ontic state λ, possibly identical to ψ in the complete case. But many instrumentalist or non-realist positions can be qualified as epistemic while not correctly taken into account by Spekkens's classification. So the possibility to avoid the conclusion that the world is not local has already been defended in previous works [14, 15].

For example, Quantum Bayesianism (QBism) mainly supported by Fuchs, Schack and Mermin [16, 17, 18, 19] is an epistemic interpretation which does not say that the wave function is about the real world. QBists think that the central subject of science is the primitive concept of experience understood as "personal experience" and that quantum mechanics is but a tool allowing any agent to compute her (subjective) probabilistic expectations for her future experience from the knowledge of the results of her past experience. They refuse the idea that the quantum state of a system is an objective property of that system. So quantum mechanics does not directly say something about the "external world". A measurement is just a special case of experience and does not reveal a pre-existing state of affairs but creates a result for the agent. In particular, there is no measurement when there is no agent: a Stern and Gerlach apparatus cannot measure by itself the spin of a particle. In this case, it becomes impossible to find an equivalent of the ontic state λ. QBists argue that they recover locality.

Similarly, Everett's interpretation [20, 21, 22] whose initial motivation is to get rid of the reduction postulate, though realist does not fit well this classification. In this interpretation, locality can also be recovered. We will describe it and discuss in details its many versions in §4.

So the above no-go theorems do not necessarily apply to these interpretations which can try to escape non-locality (what they claim).

Convivial Solipsism [23, 24, 25] which is also an interpretation without reduction postulate at the physical level is in the same case. Though accepting the existence of "entities outside of observers" it is not assumed that these entities resemble anything we are accustomed to. On the contrary, they remain in entangled states with superposed values of their properties. Their appearance is only the way they seem to be for us when we become aware of them. This, as we will show, allows to restore locality.

## 2. Is the observer playing an active role in the measurement?

The question to know what a measurement is, is one of the main problems of interpretation of the quantum formalism. Nowhere inside the formalism is defined the concept of measurement which is taken as primitive. This raises the famous measurement problem coming from the fact that there are two postulates for the evolution of a system. One is the use of the Schrödinger equation when no measurement is done on the system and the other is the reduction postulate when a measurement has been made. These two postulates do not lead to the same result but there is no clear and unambiguous way to decide when one should use one or the other (except to say that a physicist perfectly knows when a measurement is done). This is not the place here to recall the detailed way to expose the measurement problem which has been presented an incalculable number of times (see [24] for my personal presentation). Briefly stated, I argue that no satisfying solution can avoid mentioning the observer as playing a major role in the measurement. This is not something making physicists happy. Indeed, the majority of them is looking for a definition that could be regarded as "strongly objective" in the meaning that d'Espagnat gave to this term [26] (i.e. without any mention to a human observer)[1]. But all the attempts in this direction fail to provide a rigorous and clear solution. This led in the past some physicists to support the idea that the observer's consciousness was responsible for the physical collapse of the wave function [28, 29, 30]. Einstein's reply to Heitler [31] about the fact that the observer plays an important role in the collapse of the wave-function is worth considering:

*"That one conceives of the psi-function only as an incomplete description of a real state of affairs, where the incompleteness of the description is forced by the fact that observation of the state is only able to grasp part of the real factual situation. Then one can at least escape the singular conception that observation (conceived as an act of consciousness) influences the real physical state of things; the change in the psi-function through observation then does not correspond essentially to the change in a real matter of fact but rather to the alteration in our knowledge of this matter of fact."*

---

[1] See for example [27] for a typical exposition of this quest.

It is interesting to notice in this quotation the fact that observation, when understood as an act of consciousness, is seen either as implying a physical change in the psi-function or as an alteration of our knowledge. The former hypothesis is not acceptable because it comes back to a kind of dualist effect of the mind on the matter. So we are left with the latter assumption which means that the real state of things is incompletely described by the wave-function and that an observation is an updating of our knowledge about the reality. But taken at face-value, if observation is merely acquiring a more complete knowledge about the real factual situation without any change in the real physical state of things, then two successive incompatible observations should not change the real state of the system and doing again the same first observation after the second one should give the same result as the first time. But this is not true. Moreover, there is something strange in saying that *observation of the state is only able to grasp part of the real factual situation* and that *[observation corresponds] to the alteration in our knowledge of this matter of fact.* One observation can only give a partial knowledge of the real state of the system. But if the system is assumed to have a really well defined more complete state (that we cannot know completely), doing incompatible observations seems to change this state. Hence, that indicates that observation cannot be merely an update of knowledge of something that is otherwise well defined.

But there is another way to understand observation as an act of consciousness without any modification of the real state of affairs: to paraphrase Einstein's sentence, we could say: "*the change in the psi-function through observation then does not correspond essentially to the change in a real matter of fact but rather to the alteration in our perception of this matter of fact*". In this case, we do not need to assume that the wave-function is an incomplete description of a real state of affairs. Indeed, it is perfectly possible to assume that the way we perceive a given state of affairs completely described by a given wave-function depends on the observer herself and can vary when the wave-function is a superposition of eigenstates of the observable corresponding to the observation that is made. In this case, each observer perception will correspond to one of the possible eigenstates (this is a rule of Quantum Mechanics) but there is no reason why two different observers should perceive the same eigenstate. The state of the system remains nevertheless the same for the two observers and stays in a superposition. That is exactly what Convivial Solipsism states.

### 3. Convivial Solipsism: a brief overview

We begin by giving a brief overview of Convivial Solipsism that will be sufficient to compare it to Everett's interpretation and to analyse the similarities and the differences between the two interpretations. Then we will give a more detailed description. Convivial Solipsism[2] starts from the very same idea than Everett's interpretation: The dynamics of the physical world is entirely described by the Schrödinger equation and nothing else. There is no physical collapse of the wave function and the global wave function of the universe stays in a superposed state. Now, the way to explain why we, observers, see a non-superposed world is different in Everett's interpretation and in Convivial Solipsism. Contrasting these two ways is not simple since there are many different interpretations of Everett's interpretation and they do not say exactly the same things. We will focus in §4 mainly on Everett's initial presentation [20, 21], on the famous many worlds Graham and DeWitt's version [22], on Vaidman's version of the many worlds interpretation [32] and on Albert and Loewer's interpretations [33, 34].

---

[2] A detailed presentation of Convivial Solipsism and its consequences is given in [23, 24, 25].

Convivial Solipsism states that the physical world dynamics is entirely driven by the linear, deterministic dynamics of the Schrodinger equation. That is true as well of all observers who are considered as physical systems. Every interaction between two systems results in an entanglement of the systems and the wave function of the two systems becomes in general a superposed one. Again, this is true also when one of the systems is an observer who becomes entangled with the other systems. That means that the observer and her brain, considered as physical systems are entangled with the other physical systems and so, are in general in a superposed state. So we make a difference between the physical state of the observer's brain which can be superposed and the awareness felt by the observer which is always a definite result. Of course, that does not at all mean that we adopt any kind of Cartesian dualism, no more than such a dualism is adopted in Everett's interpretation where similarly, the observer is only aware of one branch whereas she is physically in a superposed state that is never reduced. What is assumed is that our brain is not equipped to make us aware of all the subtleties of a superposed state and that "the effect" that such a state makes to us is limited to what we call a definite result. It is important to emphasize that there is absolutely no effect of consciousness on matter as was assumed in the conception of Wigner and London and Bauer.

Let's consider for example a measurement of a system S in a superposed state $\Psi_S = \sum c_i |\varphi_i\rangle$, written in the basis of the eigenstates of the observable that is measured, with an apparatus A in an initial state $|A_0\rangle$ by an observer O in an initial state $|O_0\rangle$. After the interaction between the system, the apparatus and the observer, we get:

$$\Psi_{SAO} = \sum c_i |\varphi_i\rangle |A_0\rangle |O_0\rangle \rightarrow \Psi_{SAO} = \sum c_i |\varphi_i\rangle |A_i\rangle |O_i\rangle \qquad (1)$$

Where $|A_i\rangle$ is the state of the apparatus correlated to the result corresponding to the eigenvalue $\lambda_i$ of the eigenstate $|\varphi_i\rangle$ of the observable and $|O_i\rangle$ is the observer's state having seen the apparatus in the state $|A_i\rangle$.

The usual von Neumann prescription for a measurement is then to use the reduction postulate to collapse this wave function to a reduced state corresponding to one of the possible *i*, hence accounting for the fact that the observer sees only one result and not a superposition of results.

In Convivial Solipsism no such collapse occurs at the physical level and the wave function of the system, the apparatus and the observer remains entangled. How is it then possible that the observer sees a definite result? Here we must remember that the observer's perception is built inside her brain which is a limited tool. In a way, it is natural enough to assume that the human brain is not able to perceive the universe in all its subtleties. It is reasonable to assume that the observer's perception of such an entangled system is done through some kind of mental filters that prevent her to perceive the richness of the superposition and force her to see only one of the many possibilities.

Even though it is necessary to be very prudent with such a comparison, some widely popularized ambiguous pictures that it is possible to "see" in different ways can help understanding what happens. A famous one is the spinning dancer. It is bistable optical illusion[3]. Some observers initially see the figure as spinning clockwise and some counter clockwise. The question "is the dancer really spinning clockwise or counter clockwise?" has no meaning. There is no real direction of rotation. There is only a set of moving pixels that each observer interprets as a dancer spinning clockwise or counter clockwise depending on her own mental configuration at the moment. Once again with great caution, it would be possible to think that this picture is described by a superposed wave function $1/\sqrt{2}$ (|spinning clockwise⟩ + |spinning counter clockwise⟩ ). That means that the "real" state of the dancer is neither spinning clockwise nor spinning counter clockwise but instead is something that does not correspond to any classical state we are accustomed to. When we observe such a superposed state we are only able to perceive one or another of the classical states corresponding to a definite direction of rotation. Now, the fact that an observer sees the dancer spinning clockwise, has absolutely no effect on the picture which is physically unchanged. There is of course no collapse. These types of perceptive illusion (another well-known one is the Necker cube) are more and more studied by neuroscientists who are only starting to discover the brain mechanisms allowing to understand why it is so.

---

[3] See https://www.youtube.com/watch?v=9CEr2GfGilw

I mention this kind of illusion because it can help understanding what an observation is inside the framework of Convivial Solipsism. The superposed state of a system cannot be perceived as such and do not correspond to anything we are accustomed to. It is fully quantum and it remains quantum. But, when an observer gives a look at it, because of the way her brain is working she cannot see anything else than a classical state which is one of the components of the superposed state of the system. That is the reason why the observer sees a definite result while the system remains in a superposed state. In this context, a measurement is merely the fact that an observer has a look at a system (preferably a macroscopic one, this is why it is necessary to use an apparatus which correlates its states with the states of the microscopic system) and perceives what her brain allows her to perceive, that is a definite result which corresponds to a classical state. Two points are to be noticed here. The first one is that an apparatus is necessary because our senses are not suited to directly perceive microscopic things but it is in no way a fundamental need. The macroscopic nature of an apparatus plays no role in the fact that a definite result is perceived. Its only role is to allow a human observer to see something. But, had we eyes good enough, we could make a measurement of a microscopic system exactly in the same way than with an apparatus[4]. The second one is that it is pointless to wonder why classical states correspond to definite results. Asking this question assumes that there is an absolute definition of what a determinate result is, which is wrong. We have to take this question in the reverse way. It is what we are accustomed to perceive that we call definite results and classical states. So such a denomination is a posteriori and it is what we cannot directly perceive that we call superposed results. It is perfectly thinkable that individuals belonging to a differently mental oriented specie with brain differently designed perceive as "classical for them" states what we call superposed states.

Now, there are some differences between what happens in quantum mechanics and the examples of illusion I gave above. The first one is that, in these illusions, it is possible for an observer to switch from one point of view to another. With a little bit of training it is even possible to switch at will and to make the dancer spinning clockwise or counter clockwise when you want. This is of course not possible in a quantum measurement. If a spin up along the axis Oz is observed in a measurement of a half spin system by an observer, it is not possible for her to suddenly switch to a spin down if she does again the same measurement. The second one is that two observers can disagree about what they perceive about the dancer. One will say that she is seeing a clockwise movement while the other will say the contrary. This is not possible in a quantum measurement where all observers will agree on the result of a measurement. This second point can find an explanation in Convivial Solipsism as we will see, and this is precisely why this interpretation is called "Convivial" because observers can never disagree, though it is not possible to verify that they "really" perceived the same result (this is why it is called "Solipsism"). The first one is however a real difference. It is also necessary to explain why we see only certain components of the superposed wave function (those written in the so called preferred basis) and not the others. This comes from the decoherence mechanism. We will not give here the details of the explanation which is not essential for what we want to present in this paper and will refer the reader to [24] where this is explained.

---

[4] Actually, things are more complex than that. There is a subtlety due to the decoherence mechanism and the choice of the preferred basis that we forget here. See [24] for a detailed account of this point.

After this very preliminary and informal presentation of Convivial Solipsism, it is important to compare it with the different versions of Everett's interpretation and to contrast the way of explaining the definite result that an observer gets from a measurement.

### 4. Everett's interpretation

As is well-known, Everett's initial interpretation has been presented through many incompatible versions [35, 36]. As Barrett says [35]:

*"There has been considerable disagreement over the precise content of his theory and how it was supposed to work. […] There have been many mutually incompatible presentations of Everett's theory. Indeed, it is fair to say that most no-collapse interpretations of quantum mechanics have at one time or another either been directly attributed to Everett or suggested as charitable reconstructions. The most popular of these, the many worlds interpretation, is often simply attributed to Everett directly and without comment even when Everett himself never characterized his theory in terms of many worlds."*

Despite the many various versions of Everett ideas, the basis of the interpretation is very simple [20, 21]. Let's use the standard example of the measurement by an observer O of a spin ½ system S in a superposed state of z-spin with an apparatus A designed to measure the spin along Oz. If the initial state of the system is:

$$|\Psi_S\rangle = \alpha \,|+\rangle_z + \beta \,|-\rangle_z \qquad (2)$$

after the interaction between the system S and the apparatus A, the state of the big system composed of the apparatus and the spin ½ system is:

$$|\Psi_{SA}\rangle = \alpha \,|+\rangle_z |\uparrow\rangle + \beta \,|-\rangle_z |\downarrow\rangle \qquad (3)$$

where $|\uparrow\rangle$ (resp. $|\downarrow\rangle$) is the state of the apparatus correlated to z-spin + (resp. z-spin -). If an observer looks at the apparatus, then the state of the big system including the observer becomes:

$$|\Psi_{SAO}\rangle = \alpha \,|+\rangle_z |\uparrow\rangle |O_+\rangle + \beta \,|-\rangle_z |\downarrow\rangle |O_-\rangle \qquad (4)$$

where $|O_+\rangle$ (resp. $|O_-\rangle$) is the state of the observer having seen the apparatus in the state $|\uparrow\rangle$ (resp. $|\downarrow\rangle$).

Now, since Everett wants to get rid of the reduction postulate but has to explain that an observer always gets a definite result, he assumes that this superposed state must be interpreted in the following way: each element of the superposition describes an observer having seen a definite result. For justifying that, he uses what he calls the fundamental principle of relativity of states [21]:

*"There does not, in general, exist anything like a single state for one subsystem of a composite system. Subsystems do not possess states that are independent of the states of the remainder of the system, so that the subsystem states are generally correlated with one another. One can arbitrarily choose a state for one subsystem, and be led to the relative state for the remainder. Thus we are faced with a fundamental relativity of states, which is implied by the formalism of composite systems. It is meaningless to ask the absolute state of a subsystem--one can only ask the state relative to a given state of the remainder of the subsystem."*

But what does that really mean? There may be a way to understand what Everett had in mind. According to Barrett [35], for Everett, a theory is empirically faithful and hence empirically acceptable if there is a homomorphism between its model and the world as experienced: a homomorphism and not an isomorphism. That would mean that for him, there could be elements of the theory that do not directly correspond to experience. So, Everett could have been satisfied with the fact that some element of the superposition correspond to experience even if others are superfluous. It would then be possible to think that Everett thought that $|\Psi_{SAO}\rangle$ must be interpreted as containing one element corresponding to reality (which was satisfying for him) and others that needed to be neglected as superfluous. The following quotation from Barrett [35] supports this idea:

*"Note that Everett did not require a physically preferred basis to solve the determinate record problem to show that pure wave mechanics was empirically faithful. The principle of the fundamental relativity of states explicitly allows for arbitrarily specified decompositions of the absolute universal state into relative states. Given his understanding of empirical faithfulness, all Everett needed to explain a particular actual record was to show that is that there is some decomposition of the state that represents the modelled observer with the corresponding relative record."*

But actually this is not a correct interpretation of Everett's position since he said [37]:

*"Let one regard an observer as a subsystem of the composite system: observer +object-system. It is then an inescapable consequence that after the interaction has taken place there will not, generally, exist a single observer state. There will, however, be a superposition of the composite system states, each element of which contains a definite observer state and a definite relative object-system state. Furthermore, as we shall see, each of these relative object system states will be, approximately, the eigenstates of the observation corresponding to the value obtained by the observer which is described by the same element of the superposition. Thus, each element of the resulting superposition describes an observer who perceived a definite and generally different result, and to whom it appears that the object-system state has been transformed into the corresponding eigenstate."*

That means that he supported well the idea of the split of the observer in multiple copies.

Now, Graham and DeWitt give the following description [22]:

*"By virtue of the temporal development of the dynamical variables the state vector decomposes naturally into orthogonal vectors, reflecting a continual splitting of the universe into a multitude of mutually unobservable but equally real worlds, in each of which every good measurement has yielded a definite result and in most of which the familiar statistical quantum laws hold."*

So, if a measurement with a probabilistic outcome is made, the world splits into several worlds and each possible outcome of this measurement appears in some of these worlds in a proportion given by the quantum probability of that outcome.

The main difference with Everett's initial interpretation is that the universe for Graham and DeWitt is not made of a unique world while Everett claims:

*"It is ... improper to attribute any less validity or "reality" to any element of a superposition than any other element, due to [the] ever present possibility of obtaining interference effects between the elements, all elements of a superposition must be regarded as simultaneously existing"*

On the contrary, for Graham and DeWitt, there is an exponentially growing number of worlds that are causally isolated and a corresponding growing number of clones of the initial observer each one inhabiting one isolated world.

What is interesting is that it would be possible in principle to make a crucial experiment to decide between the two versions since Everett thought that all branches of the global wave function were at least in principle detectable and that, through a very complicated experiment (impossible in practice) it would be possible to observe the interferences between the branches, which is not possible in Graham and DeWitt's version. The decoherence theory, correctly interpreted, works in favour of the initial version and not of the "many-worlds" one (see [24]).

These two versions of Everett's interpretation face some well-known problems, or at least difficulties, that can bother some readers. One of these problems is to know how the preferred basis is selected. This has often been raised against the initial Everett's interpretation. Fortunately, the decoherence mechanism helps solve it and the version supported by Vaidman [32], who uses decoherence, is not subject to it. Vaidman's interpretation is close to Everett's initial one. He does not claim, as Graham and DeWitt, that there is an irreversible splitting of worlds and he even proposes an experiment to test this version of multiple worlds versus the collapse theory [38]. The main difference with Everett's initial presentation is in the way he sees the worlds. For him:

*"A world is the totality of macroscopic objects: stars, cities, people, grains of sand, etc. in a definite classically described state."*

*"Another concept, which is closer to Everett's original proposal, see Saunders 1995, is that of a relative, or perspectival world defined for every physical system and every one of its states (provided it is a state of non-zero probability): I will call it a centered world. This concept is useful when a world is centered on a perceptual state of a sentient being. In this world, all objects which the sentient being perceives have definite states, but objects that are not under observation might be in a superposition of different (classical) states. The advantage of a centered world is that a quantum phenomenon in a distant galaxy does not split it, while the advantage of the definition presented here is that we can consider a world without specifying a center, and in particular our usual language is just as useful for describing worlds that existed at times when there were no sentient beings."*

The decoherence mechanism shows that only localized states of macroscopic objects are stable. So this version escapes the criticisms made to the initial version.

Another major problem of Everett's interpretation is the way to recover probabilities. Since during a measurement all possible results are obtained, what can be the meaning of the sentence 'the probability that I will get this result is = p'? The very meaning of probability fails because each outcome should have a probability 1. The many attempts to recover probabilities are not convincing[5].

For solving the contradiction between the fact that the observer's brain can be in a superposed state while her mind is never, Albert and Loewer [33, 34] assume that every observer is associated with a nonphysical entity called "mind" which always has definite non-superposed states. In this respect, this interpretation is clearly dualist: they add something non-physical to the ontology of the theory. In the single mind version, it is assumed that after a measurement, the observer's mind chooses only one of the components of the superposed state with a probability given by the Born rule. This allows them to recover probabilities. Now, they face the so called problem of the "mindless hulks" in case (say) of a measurement of the x-spin of an electron if Alice's mind and Bob's mind are not linked to the same component. Alice would have got "up" while B would have got "down" but Alice would nevertheless hear Bob saying he got "up" and so it seems that Alice's mind would be in contact with a mindless hulk since Bob's mind inhabits the down component. This led them to postulate a continuous infinity of minds. This is the many minds interpretation. In this version, each observer is associated with an uncountable infinity of minds. During each measurement the set of minds splits in as many subsets as there are possible outcomes. This allows to populate each branch and eliminates the "empty hulks". We will come back to the single mind interpretation after having presented more precisely Convivial Solipsism in order to make clear the difference between the two interpretations which could seem similar in some respects.

Regarding the many minds interpretation, it is subject to the same criticism than the many worlds one. As Barrett states [41]:

"*The difference between a many-worlds theory […] and a many-minds theory (like AL's) is not that great. It really just amounts to a difference in what sort of facts one takes as being determinate. On a many-worlds theory physical records and thus local mental states are determinate, and on a many-minds theory it is only local mental states that are determinate. If one had a successful many-minds theory, then one could always convert it to a many-worlds theory.*"

First, it is true that this stupendous proliferation of worlds (or of minds) is something that one can find repulsive even though it could be considered as a simple matter of taste. There is nevertheless something not very clear. On the one hand, one may speak of different observers described by different states. On the other hand, the same physical system is involved, and from this viewpoint it is the same observer, which is in different states corresponding to different elements of the superposition. So from an external point of view there is always only one observer. What then does that mean that there are several observers? For whom are there several observers? It is not true that for the observer involved in the measurement, there are other observers corresponding to herself since she cannot be in contact with them. For an external observer not involved in the interaction having caused the entanglement of the big wave function, there is only one system and one entangled observer having made a measurement.

---

[5] See for example Wallace [39] who claims having proved the Born rule in this context and Kent [40] who denies that it is the case.

Hence, from either point of view (internal or external) there is always one world! The problem is that the story of many worlds (or of many minds) is told from the point of view of an observer being able to witness all the different observers, which is not possible.

On the top of that, one feels well that there is a tension between classical and quantum in these "many worlds -minds-" interpretations, as can be seen in the description given by Vaidman of a centered world [32]: *"In this world, all objects which the sentient being perceives have definite states, but objects that are not under observation might be in a superposition of different (classical) states. The advantage of a centered world is that a quantum phenomenon in a distant galaxy does not split it."* According to this description, within this type of world there is a mix between a part that is classical (what is observed) and a part that is quantum (what is not observed). This shows clearly that there is a kind of malaise forcing the author to welcome a mixture of classical and quantum which seems very artificial: in these kind of worlds there are classical parts and quantum parts, which, for this world, are both real.

This raises a criticism that I find the most serious: the fact that all versions of Everett's interpretation are somehow prisoner of a classical way of thinking. Why should one consider that a system in a superposed state $1/\sqrt{2}(S_1 + S_2)$ of two classical states $S_1$ and $S_2$ gives rise actually to two systems, the first one in the classical state $S_1$ and the second one in the classical state $S_2$ as soon as an observer has a look at it? Just because it seems to the observer that she sees $S_1$ or $S_2$? I do not think that it is a good reason. It should be acknowledged that the world can really be (and stay) quantum and only quantum and that, in this case, the system stays in the superposed state $1/\sqrt{2}(S_1 + S_2)$. This is actually what Everett assumes at the beginning but he does not draw all the consequences of this assumption. There is no need to assume the birth of many observers or of many worlds after the measurement. We just have to account for what the observer perceives from this superposed physical reality. Actually, what we call classical is everything we have been accustomed to through our usual experience so it is not surprising that the world that we perceive is classical even if the world itself is quantum. So we should not accept an interpretation which relies on a decomposition of the wave function done in a way that voluntarily selects classical components and then declares that the world is identical to the juxtaposition of these components supposed to exist simultaneously (with the possibility to interact or not depending on the fact that the initial version or the version from Graham and DeWitt is adopted). Claiming that a system in a superposed state of two classical states is in fact equivalent to two systems each one in one of the classical states is not acceptable, no more than claiming that this superposition represents two different classical worlds. Yet this amounts roughly to the position of all the versions I have presented above. On the contrary, we must clearly make a demarcation between the world and the perception we have of it and so, acknowledge the fact that the world is fully quantum, that the state of the system is "really" $1/\sqrt{2}(S_1 + S_2)$ and that we do not need classical worlds except for describing the way we can perceive the world. So doing, we do not need many worlds (or many minds) anymore.

In this respect I agree with Vaidman [42]:

*"I find by far the best option to take the wave function of the Universe, and only it, as the ontology of the theory"*

Even if we will see that, due to the relativity of states, the ontology of Convivial Solipsism is more complex than that because there is no unique absolute wave function of the universe, this point is fundamental for Convivial Solipsism. It shares with Everett's interpretation the rejection of the collapse but tries to keep only one world and one observer.

## 5. Convivial Solipsism: formal presentation

Convivial Solipsism starts from exactly the same will to cancel the collapse of the wave function. The initial motivation for Convivial Solipsism comes from a remark of d'Espagnat [43] who was totally unsatisfied by the astronomical number of worlds (or of minds) that is assumed in the various presentations of Everett's interpretation. Convivial Solipsism is an interpretation inside which the physical dynamics of the universe is described by the Schrödinger equation and which states that a measurement is nothing else that the awareness of a result by a conscious observer. A superposed wave function cannot be perceived by a human observer in all its richness and consequently, an observer watching such a superposed physical thing will perceive it through some mental filters allowing her to see only a definite result and giving her the impression that a definite result is "really" obtained and that the system is "really" in the state she perceives. But actually, what she sees is just the result of the limitation of her perception by these filters and the system remains in a superposed state. This awareness has no physical impact on the systems that are measured and whose states are unchanged after the measurement. The reduction is only a way to describe what appears to the observer and does not affect the "reality" which remains superposed. Hence no physical collapse is necessary. Convivial Solipsism rests on two main assumptions.

### 5.1. First assumption: the hanging-on mechanism

*"A measurement is the awareness of a result by a conscious observer whose consciousness selects at random (according to the Born rule) one branch of the entangled state vector written in the preferred basis and hangs-on to it. Once the consciousness is hung-on to one branch, it will hang-on only to branches that are daughters of this branch for all the following observations."*

The last part of the hanging-on mechanism[6] guarantees that repeating the same measurement will give again the same result. Let's take again the example of the measurement of a spin 1/2 particle in a superposed state along Oz.

$$|\Psi_S\rangle = \alpha \, |+\rangle_z + \beta \, |-\rangle_z \tag{5}$$

After the interaction with the apparatus the global state is:

$$|\Psi_{SA}\rangle = \alpha \, |+\rangle_z |\uparrow\rangle + \beta \, |-\rangle_z |\downarrow\rangle \tag{6}$$

If we include now the physical state of an observer having a look at the apparatus, we get:

$$|\Psi_{SAO}\rangle = \alpha \, |+\rangle_z |\uparrow\rangle |☺\rangle + \beta \, |-\rangle_z |\downarrow\rangle |☹\rangle \tag{7}$$

where $|\uparrow\rangle$ (resp. $|\downarrow\rangle$) is the state of the apparatus correlated to spin + (resp. spin -) and $|☺\rangle$ (resp. $|☹\rangle$) is the physical state of the observer's brain correlated to the state $|\uparrow\rangle$ (resp. $|\downarrow\rangle$) of the apparatus.

As noticed before, we make a difference between the physical brain of the observer and her awareness. $|☺\rangle$ and $|☹\rangle$ are physical states of the observer's brain. Now, the hanging-on mechanism says that the consciousness of the observer chooses one branch at random. Let's

---

[6] Actually the hanging-on mechanism is a little bit more complex if both decoherence and the relativity of states (see below) are taken into account, but this has no impact on what we want to say here. See [24] for a more detailed presentation.

denote by ☺̃ the fact to be aware of having seen "+" (resp. by ☹̃ of having seen "-"). After the hanging-on mechanism, it will be either the case that the observer has seen "+" or that the observer has seen "-". So after the measurement, the wave function of the global system is still described by (7) but the observer's awareness is either ☺̃ or ☹̃. We must be very clear not to confuse $|☺\rangle$ with ☺̃ and $|☹\rangle$ with ☹̃. $|☺\rangle$ and $|☹\rangle$ are kets describing the physical states of the observer's brain that enter into the entangled universal sate vector. ☺̃ and ☹̃ are not state vectors and cannot enter into any linear combination with state vectors. That is the reason why they are not written as kets. They are just describing states of awareness.

Now, according to the hanging-on mechanism, once the consciousness is hung-on to one branch, it will stick to it and will hang-on only to branches that are daughters of this branch for all the following observations. So even though the wave function is unchanged and remains the superposed (7), for all subsequent measurements everything happens for the observer as if the wave function was reduced either to $|+\rangle_z |\uparrow\rangle |☺\rangle$ if her state of awareness is ☺̃ or to $|-\rangle_z |\downarrow\rangle |☹\rangle$ if her state of awareness is ☹̃. This insures that repeating the same measurement will give the same result.

Of course, contrary to Everett's interpretation, there is no problem of accounting for probabilities in Convivial Solipsism since the Born rule is explicitly assumed as it is the case in standard quantum mechanics. But the incoherence of having two different rules without having a clear prescription for when the reduction postulate must be used (because there is no clear definition of what a measurement is), is fixed in Convivial Solipsism. Convivial Solipsism solves the problem that worried Everett without falling into the trap of having to explain where probabilities come from. It is perhaps necessary to make this point clear. Convivial Solipsism does not claim that probabilities spontaneously emerge from its formalism. It just uses the standard Born rule exactly as it is used in standard Quantum Mechanics with collapse. But the problem of Quantum Mechanics with collapse is not the fact that Born rule is assumed: nobody says that probabilities must absolutely be emerging from the formalism. The problem in standard Quantum Mechanics is that it is impossible to define unambiguously what a measurement is and when the Born rule should be used. So the Born rule is not problematic by itself but the conditions of its use are unclear. This is the notorious measurement problem. Convivial Solipsism has been built first for solving the measurement problem. When this is done, it becomes possible to use the Born rule in a rigorous way and to know clearly what probabilities are about: probabilities are about the states of awareness because between many possible ones, only one appears. But contrary to what has been sometimes said, it is not enough to impose by fiat the Born rule to solve the problem in an interpretation which does not make the meaning of probabilities clear. Since in Everett's interpretation all possibilities happen, the very concept of probability fails. As already mentioned above, the problem is of making sense of statements like 'the probability that I will get this result is = p'. Simply assuming the Born rule is of no help to solve this problem. This is the reason why supporters of Everett's interpretation try (unconvincingly from my point of view) to make probabilities emerging from the formalism because, in the interpretation taken at face value, there is no room for any probability.

It is then possible to paraphrase Vaidman [32] replacing MWI by Convivial Solipsism:

*"Convivial Solipsism is a deterministic theory for a physical Universe and it explains why a world appears to be indeterministic for human observers."*

God does not play dice but we, observers, do.

### 5.2. Second assumption: the relativity of states

All what has been said until now, could be understood as if there was a unique independent real world that all the observers witness (even if they are part of it) and which is described by the entangled global wave function that is never reduced. Of course, each observer would have her own private conscious experience of this universe (that could be different from one observer to the other depending on the branch to which she is hung-on, this is why it is a kind of soft solipsism) but the global wave function would be the same for all the observers and would be "The" wave function of a universe conceived as unique and independent of the observers. That is the point of view of Everett's interpretation. In this case, we would recover a certain kind of standard realism even if the universe so described would be very different from the universe that each observer perceives. But Convivial Solipsism shares with QBism [16, 17, 18, 19] and Rovelli's relational interpretation [44, 45, 46] the idea that the entities inside the quantum formalism (state vectors and probabilities) are relative (to the agent for QBism, to the system for Rovelli and to the observer in Convivial Solipsism).

A major common point between Convivial Solipsism and QBism is the fundamental role that the observer plays in the measurement process and the relinquishment of any absolute description of the world. Now, QBists refuse the idea that the quantum state of a system is an objective property of that system. It is only a tool for assigning a subjective probability to the agent's future experience. So quantum mechanics does not directly say something about the "external world".

*"But quantum mechanics itself does not deal directly with the objective world; it deals with the experiences of that objective world that belong to whatever particular agent is making use of the quantum theory."* [19].

On the contrary, Convivial Solipsism[7] states that quantum mechanics says something about the external world of each observer and the way this external world together with the mental equipment of the observer makes the observer's perception be what it is.

The second assumption is then:

*"Any state vector is relative to a given conscious observer and cannot be considered in an absolute way."*

In Convivial Solipsism, the state vectors, including the universal entangled one, are relative to one observer. That means that even the universal entangled wave function has no absolute and universal validity but is relative to each observer. It represents only the description of the universe for a specific observer. In this case, one may ask what that means to continue speaking of a universe. The universe is no more an absolute reality existing outside and independently of any observer but is relative to each observer. That does not mean however that nothing else than the mind exists and that the universe is totally created by the consciousness of each observer. That would amount to coming back to a pure idealist position which is not what

---

[7] See [24] for a more detailed comparison between Convivial Solipsism and the relational interpretation and QBism and for a description of the issues that, from my point of view, these two latter interpretations face.

Convivial Solipsism lauds. Indeed, it is well known that pure idealism faces many difficulties and that strict solipsism, while not refutable, is not a very interesting position. As I explain in [24]:

*"Convivial Solipsism is situated in a neo-Kantian framework and assumes that there is "something" else than consciousness, something that (according to the famous Wittgenstein's sentence) it is not appropriate to talk of. This is close to what Kant calls "thing in itself" or "noumenal world". Consciousness and this "something" give rise to what each observer thinks is her reality, following Putnam's famous statement "the mind and the world jointly make up the mind and the world". So perception is not a passive affair: perceiving is not simply witnessing what is in front of us but is creating (independently for each us) what we perceive through a co-construction from the world and the mind. The hanging-on mechanism takes part in this co-construction and helps (very partially) understanding it through the selection it does."*[8]

We can now analyse the differences between Convivial Solipsism and the single mind interpretation. First, as Albert and Loewer say themselves, the single mind interpretation is dualist, which Convivial solipsism is not. There are no non-physical minds added to the ontology. Second, the picture given by the single mind interpretation, like the other versions of Everett's interpretation, is a picture of co-existing juxtaposed classical worlds (able or not to communicate between them) corresponding to all possible results. So, as soon as we imagine that one of these worlds is inhabited by the unique mind of the observer, all the other worlds are mindless and bodies living in these worlds are zombies. Let's take again the case (say) of a measurement of the x-spin of an electron if Alice's mind and Bob's mind are not inhabiting the same world. Alice would have got "up" while B would have got "down" but Alice would nevertheless hear Bob saying he got "up" and so it seems that Alice's mind would be in contact with a mindless hulk since Bob's mind inhabits the down component.

Nothing similar happens in Convivial Solipsism. There is no juxtaposition of worlds inhabited or not by minds. The image is more radical. For each observer there is only one world, and this world stays quantum (hence superposed) and is described by a state which is relative to this observer. Inside a world, everything except the observer from whom the point of view is adopted, is considered as a physical system. This includes the other observers who, for her, are mere physical systems having a status similar to any other system like an electron or an apparatus. There is no mind talking to other minds. There is only an observer equipped with her own wave function describing her universe (herself included) who makes observations on physical systems and becomes aware of results according to the hanging-on mechanism allowing her to witness only a part of the superposed state. In this framework, it is meaningless to wonder if when Alice hears Bob saying he got down, he really got down. No observer has a direct access to the perceptions of another observer. Then, it is not allowed from Alice's point of view to speak of what Bob's real awareness is. From Alice's point of view, Bob is nothing more than a physical system and hearing Bob saying that he got "down" has to be treated exactly on the same foot that seeing the needle of a Stern and Gerlach apparatus pointing downside, nothing more. Remind that in Convivial Solipsism this is not because an observer sees that the needle points down that it "really" points down. The needle stays in a superposed state but the observer's perception selects the down position. This is the same thing that happens when Alice speaks with Bob to know what he saw. As detailed below in 5.3, Bob stays in a superposed state but Alice's perception selects one among Bob's possible answers. Moreover, it is possible that in Bob's universe, he never made any measurement of the spin of the electron Alice measured even if in Alice's world, it seems to Alice that he did. Of course that does not mean that Alice has no right to wonder if Bob really saw "down" and the answer is that everything goes for her as if it was the case. Nevertheless talking about what Alice and Bob really saw is meaningless and forbidden because a sentence saying "Alice saw this and Bob saw that" would have to be enunciated by a meta-observer, able to witness directly the states of awareness of Alice and Bob. Now such a meta-observer does not exist.

---

[8] Of course, we do not pretend being able to explain how awareness happens. This is probably one of the most difficult problems of the contemporary science. Neuroscientists are hardly beginning to understand some mechanisms showing how consciousness works and how different it is from what our own consciousness itself thinks it is working.

### 5.3. Why is Convivial Solipsism, convivial?

Since the hanging-on mechanism is relative to each observer, a natural question is to know if there can be any conflict between two observers, the first one having seen a result and the second one having seen something different. The answer is that this is not possible. Very similarly to what happens in QBism, the communication between two observers A and B is considered as a measurement. As d'Espagnat [43] puts it:

*"Any transfer of information from B to A—for example, any answer made by B to a question asked by A— unavoidably proceeds through physical means. Therefore it necessarily takes the form of a measurement made by A on B. And we know that under these conditions A necessarily gets a response (answer) that agrees with his own perception".*

As we said before, for observer A, observer B is nothing else than a physical system. Hence, when an observer, say Alice, speaks with another one, Bob, it is as if Alice was doing a measurement on Bob. Assume that Alice and Bob do a measurement along the axis Oz of the spin of an electron in an initial superposed state of spin. From Bob's point of view, a measurement has been made so he knows the value of the spin of this electron. According to the hanging-on mechanism Bob's consciousness is hung-on to one of the two possible branches linked with the results "up" or "down". Now from Alice's point of view, Bob is entangled with the electron, as described by the Schrödinger equation. Alice having done also the measurement of the spin along Oz, her consciousness is hung-on as well to one of the two branches. The crucial point is that this branch includes the state of Bob that is linked to the very same value. So when Alice, hung-on to that branch, speaks with Bob to know what Bob saw, she performs a measurement on Bob and, accordingly to the hanging-on mechanism, she cannot hear Bob saying anything else than the value that she has got herself. Alice will never hear Bob saying that he saw "up" when she saw "down". No conflict is possible and the intersubjectivity is preserved. So, we are led to a sort of solipsism because the consciousness of each observer is located inside its own branch of its own relative universal state vector independently of the others (in a sense that is true also of each observer in Everett's picture). But that is not a true solipsism as it welcomes both other minds and an external stuff that is independent of the mind. It is closer to Kant transcendental idealism. Now, it is convivial since no conflict is possible: the quantum rules together with the hanging-on mechanism for each observer prevent any possibility to notice a divergence between the perceptions of two different observers.

The process of communication between two observers that we have described is very important because we are going to see that it is the key allowing to understand how Convivial Solipsism avoids non-locality.

## 6. Convivial Solipsism: locality or non-locality?

Let's briefly recall the way non locality is usually inferred from EPR paradox and Bell's inequalities. We give here the standard presentation of EPR paradox and a simple demonstration of Bell's inequalities.

### 6.1. The EPR paradox

The initial aim of Einstein, Podolski, Rosen paper [47] was to prove that quantum mechanics is not complete. They first state the now famous Criterion of Reality:

"*If, without in any way disturbing a system, we can predict with certainty (i.e., with probability equal to unity) the value of a physical quantity, then there exists an element of reality corresponding to that quantity.*"

Then they propose a thought experiment in which two quantum systems interact in such a way that two conservation laws hold, the relative position along the x-axis and the total linear momentum along that same axis which is always zero. They then say that measuring the position of the first system allows to predict with probability one the position of the second one. According to the Criterion of Reality, that means that the predicted position is an element of reality for the second system. The same reasoning is possible if this is the momentum of the first system that is measured. Hence, both position and momentum are elements of reality for the second system. This is in contradiction with quantum mechanics that claims, if it is complete, that there can be no simultaneously real values for incompatible quantities. To be correct, this demonstration needs nevertheless two additional postulates. The first one is separability: at the time when a measurement is made on the first system, the second system maintains its separate identity even though it is correlated with the first one. The second postulate is locality: at the time of measurement, the two systems no longer interact so no real change can take place in the second system as a consequence of a measurement made on the first one. As Fine puts it [48]:

"*In summary, the argument of EPR shows that if interacting systems satisfy separability and locality, then the description of systems provided by state vectors is not complete. This conclusion rests on a common interpretive principle, state vector reduction, and on the Criterion of Reality.*"

Of course, the EPR argument was widely discussed at the time in particular by Bohr. In substance, Bohr acknowledges the fact that the measurement of the first system does not involve any mechanical disturbance of the second system. But he claims that this measurement on the first system does involve "an influence on the very conditions which define the possible types of predictions regarding the future behavior of the other system." Now measuring the position of the first system does not allow any prediction for the momentum of the second system. Since for Bohr the very meaning of a property depends on the experimental set up that is used, it is not possible to speak of the momentum as being an element of reality of the second system if the position is what is measured on the first one. Hence for him, the argument does not hold anymore. Since it is no more possible to consider that both observables are defined simultaneously, position and momentum cannot have definite values at the same time and quantum mechanics can again be considered as complete. A careful analysis shows nevertheless that if Bohr succeeds in refuting the incompleteness of quantum mechanics this is at the price to giving up either separability or locality. Indeed, claiming that it becomes possible to speak of the position of the second system only as soon as the position of the first one has been measured, as far as the two systems can be from each other, means either that the two systems are actually one unique system before this measurement or that the first one can have a sort of instantaneous action at a distance on the second one.

It has become usual to state the EPR paradox not through the initial formulation with position and momentum as incompatible observables but through the formulation given by Bohm [49]. In Bohm's formulation, a spin zero particles decays with spin conservation into two spin ½ particles U and V in the singlet state of spin. Such a state can be written as:

$$|\psi\rangle = \frac{1}{\sqrt{2}}\left[|+\rangle^U |-\rangle^V - |-\rangle^U |+\rangle^V\right] \qquad (8)$$

Where $|+\rangle$ and $|-\rangle$ are the states corresponding to a spin +1/2 and -1/2 of the related particles along an arbitrary chosen axis. The main point is that this state is invariant by rotation and that the total spin of the two particles along any axis must be zero. Hence, if a measurement of the spin of the particle U along one arbitrary axis is "+" then a measure of the spin of the particle V along the same axis will have to give "-". Now the same reasoning as the previous one with position and momentum in the initial formulation can be done. In this case, the incompatible observables are for example, the spin along x-axis and the spin along z-axis. Briefly stated, if a measurement of the spin along x-axis of U is made, this allows to predict with probability one the value of the spin along x-axis of V. Hence, according to EPR Criterion of Reality, the value of the spin along x-axis of V is an element of reality for V. As this can be done also for the measurement of the spin along z-axis of U, that means that the value of the spin along z-axis of V is also an element of reality for V. Hence, both the value of the spin along x-axis and the value of the spin along z-axis of V are simultaneous elements of reality for V. But, this is in contradiction with the fact that quantum mechanics is complete since according to the quantum formalism, the observable spin along x-axis and the observable spin along z-axis are not commuting and cannot have simultaneously well-defined values.

Let's first make clear that the situation is not similar to the classical situation where "things are done" when the two particles separate. Assuming that, would mean that the explanation for the fact that the value of the spin along z-axis of V is "-" (resp. "+") if the measurement of the spin along z-axis of U have given the result "+" (resp. "-") is simply the fact that as soon as U and V separate, the value of the spin along z-axis of U was already "+" (resp. "-") and the value of the spin along z-axis of V was already "-" (resp. "+"). That would mean that the state of U and V right after their separation is either $|\psi_1\rangle = |+\rangle^U |-\rangle^V$ or $|\psi_2\rangle = |-\rangle^U |+\rangle^V$. In this case, if we consider a set of N pairs U and V, that means that this set is a mixture of N/2 pairs in the state $|\psi_1\rangle$ and N/2 pairs in the state $|\psi_2\rangle$. But this is in contradiction with the fact that the pairs are in the singlet state. Indeed, the predictions given by N systems in the singlet state and those given by this mixture are different as soon as another axis is considered. For example, the probability for finding the same spin "+" along x-axis for U and V is ¼ in the case of the mixture while it is 0 in the case of the singlet state. This proves that it is not possible to consider that the value of the spin along any axis is already fixed as soon as U and V separate and before any measurement. It is only when a measurement is done on U along one particular axis that the value of the spin along this axis becomes defined.

Now comes the striking point: if this is true, that means that the value of the spin of V along the same axis is also only determined when the measurement on U is done, whatever the distance between U and V be. So a measurement of the spin of U along an axis R providing the result "+" has the effect of projecting the singlet state vector of the pair into a new state that is a tensorial product of two pure states corresponding to a definite value of the spin along this axis:

$$|\psi\rangle = \frac{1}{\sqrt{2}}\left[|+\rangle^U |-\rangle^V - |-\rangle^U |+\rangle^V\right] \xrightarrow{becomes} |+\rangle_R^U |-\rangle_R^V \qquad (9)$$

So, the state vector of V becomes $|-\rangle_R^V$ immediately after the measurement on U having given the result "+". Hence, this measurement has three effects. It determines the value of the spin of U along the axis R, it separates U and V allowing each of them to possess its own state (which was not the case before because they were entangled) and it determines the value of the spin of V along the axis R. This measurement has then an influence on the state of V and this influence at a distance is instantaneous. This conclusion is embarrassing especially for those realists who assume that the state vector is representing a real state of the system, for a change in the state vector is, for them, a change in the real state of the system. Hence it seems that locality is violated in some respects even if no mechanical disturbance is involved and even if it can be shown that is not possible to transmit any information through this process.

To be clear, let's summarize the reasoning at this stage: either quantum mechanics is not complete because things are already determined before the measurement (which is a situation that it is not possible to describe inside the quantum formalism and that leads to assuming hidden variables) or there is a violation of locality. This violation is more or less serious depending on the level one is assuming a realist position, whether one interprets the state vector as representing a real state of the system or not and whether one considers the collapse as a real change in the physical state of the system or not.

The EPR argument was only a thought experiment until Bell came.

### 6.2. Bell's inequalities

Let's call a "local theory" a theory where the outcome of an experiment on a system is independent of the actions performed on a different system that has no causal connection with the first. Let's define as realist a theory where it is meaningful to assign a property to a system independently of whether the measurement of this property is carried out. Bell shows [1] that, under the assumption that a theory is local and realist, certain correlations that can be measured between the two systems in a EPR experiment must obey some inequalities. The interesting point is that quantum theory predicts that Bell's inequalities must be violated. That means that quantum theory is inconsistent with the assumptions used to derive the inequalities. Then quantum mechanics is either non-local or non-realist.

The initial form of Bell's inequality was not convenient for setting up an experiment to test if the inequality is actually violated by quantum mechanics. Many different variants of Bell's inequality have been derived. The most usual form for the experiments is the Bell-Clauser-Horne-Shimony-Holt (BCHSH) inequality [50] and it concerns the polarization of photons. As is well known, even if it is always possible to question one or another subtle detail in each experiment to escape the conclusion, it is now widely agreed, after Aspect's first experiments and the many more recent loop-hole free tests [7, 8, 9, 10, 11, 12, 13], that the inequalities are violated and that the result given by quantum mechanics is the correct one.

What does that mean that no theory can respect both realism and locality? It is well known that the quantum mechanics formalism provides no way to describe the fact that incompatible observables (such as position and momentum or spin along two different axes) have simultaneously definite values. In this sense, quantum mechanics is not directly compatible with realism (at least understood according to the way presented above) and this is the reason why Einstein thought that it was not complete. But, it could be possible to imagine that it is possible to complete it with hidden variables which describe the "real" values possessed by the incompatible observables. That's what Bohm's theory does. But of course, in order to be empirically correct, this theory must give the same predictions as ordinary quantum mechanics (and this is actually the case). So Bell's inequalities are also violated in Bohm's theory which means that locality is not respected (since realism seems to be, even if in a very peculiar way). More generally, any hidden variables theory must be non-local. That means that even if incompatible properties can have simultaneously definite values independently of any measurement, any measurement of one observable on one system can possibly affect instantaneously the value of the corresponding observable on another distant entangled system. If we summarize, the EPR argument shows that either quantum mechanics is not complete or it is not local. Bell's inequalities show that any hidden variables theory must be non-local. Hence, non-locality seems unavoidable.

We are nevertheless going to see that Convivial Solipsism allows quantum mechanics to be local even if it is at the price of reinterpreting the formalism in a rather radical way.

### 6.3. Convivial Solipsism and locality

The analysis of EPR paradox shows that if we want to stay inside the framework of standard quantum mechanics and that we think it to be complete, it is not possible to consider that the value of the spin along any axis is already fixed as soon as U and V separate and before any measurement. It is only when a measurement is done on U along one particular axis that the value of the spin along this axis becomes defined. The value of the spin of V along the same axis is also only determined (as the opposite of the value of U) when the measurement on U is done, whatever the distance between U and V. This measurement seems then to have an instantaneous influence at a distance on the state of V. Now, if the state vector represents a real state of the system, a change in the state vector is a change in the real state of the system. This is the non-locality.

Let's analyse more precisely how this conclusion is obtained. Imagine Alice measures the spin along Oz of U and Bob the spin along Oz of V. Imagine Alice makes her measurement first. The only way for Alice to know the value of the spin of V measured by Bob afterwards is to ask him. If she does, she will invariably get an answer opposite to the result she got, in agreement with the fact that the total spin is null. Then, as she knows that before the two measurements, neither the spin of U nor the spin of V was defined, she will think that her measurement on U caused the spin of V to become defined and that Bob's measurement only noticed what the value that her measurement instantaneously produced was. In her mind, as soon as she performed her measurement, the value of the two spins became instantaneously determined. Indeed Bob could have done his measurement immediately after hers even if the distance between U and V was very large and he would have recorded the same result. The effect of Alice's measurement is then non-local since it produces an effect instantaneously whatever the distance between U and V. The same reasoning can be made in reverse if Bob's measurement is the first and can then be considered as causing the spin of U to become determined.

Now even if the two measurements are space-like separated, the two results will invariably be opposite so that the total spin is null. But as in this case, it is impossible to say in an absolute way which measurement has been made first, it becomes problematic to say that one of them is the cause of both results. This problem is hardly noticed in the usual presentations of EPR paradox which do not hesitate to accept the fact that one measurement causes the collapse of the entangled wave function and hence the determination of both spins. It is often considered enough to notice that even if the collapse is instantaneous, it cannot be used to transmit information and then does not violate relativity. But this issue casts doubt on the fact that one measurement on one of the two particles can be the cause of the determination of the value for the other[9].

Let's examine more carefully the way Alice draws the conclusion that her measurement on U caused instantaneously the determination of the spin of V. Imagine that Bob's measurement is time-like separated from Alice's one which is the first one. It is clear that this is not from a direct observation of the spin of V immediately after she saw the spin of U that Alice knows the spin of V. She knows it only after having communicated with Bob through a way which necessarily respects the speed of light limitation. Of course, in the case where Bob's measurement is made at a time allowing a standard action at a distance causing the determination of the result he gets from the measurement made by Alice, there is no need of non-locality. But the point is that Alice can think, in a counterfactual way, that even if Bob had made his measurement well before, at a time not allowing a standard lower than speed of light action, he would nevertheless have got the same result[10]. Hence, this is a proof for her that the spin of V was determined as soon as she made her measurement on U whatever the distance between U and V be. Of course, this is a reasonable deduction in a realist framework where there is a unique reality which is the same for all the observers. But in Convivial Solipsism, the way to interpret these results is different. Assume that Bob's and Alice's measurements be space-like separated and that Alice gets "+". Later Alice will get from Bob the result of his measurement and it will be "-". But the counterfactual reasoning allowing her to infer that this means that the spin of V has really been "-" immediately after the moment she made her measurement (or after the moment Bob made his measurement, since in this case it is not possible to determine which one was the first) is no longer correct. What Convivial Solipsism states is that, in agreement with the hanging-on mechanism, when Alice made her measurement, her awareness hung-on to the branch "+" of the entangled wave function. But nothing happened at the physical level neither to U nor to V. Remember that for Alice, Bob is nothing else than a physical system and he is entangled with U and V and also in a superposed state. Asking Bob about his result is equivalent to measuring Bob. When she does that, in her future light cone, the hanging-on mechanism says that Alice can only get a result given by the branch she is hung-on. That means that she can get nothing else that "-". But that does not mean anymore that the physical state of V has been "-" as soon as she did her first measurement. Actually, the physical state of U, V and Bob remains entangled and does not change. Alice's measurement is only the fact that her awareness hung-on to one branch of the entangled global wave function while there is no physical change. There is no non-locality anymore.

---

[9] See [25] for a more detailed discussion of this point.
[10] Actually, experiments showing the violation of Bell's inequalities in such a situation have been carried on.

## 7. Conclusion

While aiming at the same goal than Everett's interpretation to get rid of the physical collapse of the wave function and presenting also a picture of the world which is deterministic and local, Convivial Solipsism avoids some of its difficulties. A first point is to avoid this tremendous proliferation of worlds. A second point is the fact that, so doing, it becomes possible to reintroduce probabilities much in the same way as in standard Quantum Mechanics but without being faced to the measurement problem. A third point, as discussed above, is the fact that Everett's interpretation seems a prisoner of a classical way of thinking which is not the case with Convivial Solipsism.

Convivial Solipsism shares also some similarities with QBism, another non-collapse approach. But, it is much clearer in the picture it gives and avoids many unclear points in QBism: the way an agent impinges on the world, how it is possible for an agent to create a result and more important, the ambiguity of what is necessary for a system to be an "agent", which is left unclear in QBism. These points are discussed more thoroughly in [24].

It is true that this is obtained at the price of modifying the conception we have of reality, which is not longer the same for all the observers and which is fundamentally different from what we perceive. The picture given by Convivial Solipsism can be considered very weird. But abandoning the intuitive image that we got from our classical experience is now widely accepted as mandatory. This is more or less already the case, while in many different ways, in all the attempts to interpret quantum mechanics. Moreover the description given by Convivial Solipsism of the articulation between the "world" and the perceptions that an observer has of it seems in agreement with current trends in neuroscience. Convivial Solipsism is logically coherent and solves many questions that remains puzzling in some other interpretations.

## Acknowledgement

I want to thank Chris Fuchs for many enlightening discussions about QBism and Lev Vaidman for exchanges allowing me to better understand his own presentation of Everett's interpretation.